\documentclass{cidr-2019}

\usepackage{enumitem}
\usepackage{framed}
\usepackage{cprotect}
\usepackage{enumitem}
\usepackage{listings}
\usepackage{amstext}
\usepackage{amstext}
\usepackage{pdfpages}
\usepackage{alltt}
\usepackage{epstopdf}
\usepackage{xspace,colortbl}
\usepackage[USenglish]{babel}
\usepackage{multirow}
\usepackage{subfigure}
\usepackage{graphicx}%%
\usepackage{amssymb}
\usepackage{fmtcount}
\usepackage{amsfonts}
\usepackage{xspace}
\usepackage{amsmath}
\usepackage{multirow}
\usepackage[mathscr]{eucal}
\usepackage{colortbl}

\usepackage{amsmath,amssymb}
\usepackage[linesnumbered, ruled,vlined]{algorithm2e}

\usepackage{caption}
\usepackage{graphicx}

\usepackage{bm}
\usepackage[nospace]{cite}
\usepackage{csquotes}
\usepackage{enumitem}
\usepackage{charter}

\usepackage{cleveref}

\begin{document}

\newcommand{\detectlib}{\texttt{IsoDetect}\xspace}
\newcommand{\company}{\texttt{Company X}\xspace}
\newcommand{\cond}{\textrm{pred}\xspace}
\newcommand{\dataset}{data set\xspace}
\newcommand{\datasets}{data sets\xspace}
\newcommand{\spview}{\textsf{SPView}\xspace}
\newcommand{\fjview}{\textsf{FJView}\xspace}
\newcommand{\aggview}{\textsf{AggView}\xspace}
\newcommand{\hashfunc}[1]{\textsf{hash}(#1)\xspace}
\newcommand{\hashop}{\textsf{hash}\xspace}
\newcommand{\nsc}{\textsf{NormalizedSC}\xspace}
\newcommand{\rsc}{\textsf{RawSC}\xspace}

\newcommand{\avgfunc}{\ensuremath{\texttt{avg} }\xspace}
\newcommand{\maxfunc}{\ensuremath{\texttt{max} }\xspace}
\newcommand{\minfunc}{\ensuremath{\texttt{min} }\xspace}
\newcommand{\histfunc}{\ensuremath{\texttt{histogram\_numeric} }\xspace}
\newcommand{\countfunc}{\ensuremath{\texttt{count}}\xspace}
\newcommand{\sumfunc}{\ensuremath{\texttt{sum} }\xspace}
\newcommand{\varfunc}{\ensuremath{\texttt{var} }\xspace}
\newcommand{\stdfunc}{\ensuremath{\texttt{std} }\xspace}
\newcommand{\covfunc}{\ensuremath{\texttt{cov} }\xspace}
\newcommand{\corrfunc}{\ensuremath{\texttt{corr} }\xspace}
\newcommand{\medfunc}{\ensuremath{\texttt{median} }\xspace}
\newcommand{\percfunc}{\ensuremath{\texttt{percentile} }\xspace}
\newcommand{\havingfunc}{\ensuremath{\texttt{HAVING} }\xspace}
\newcommand{\selectfunc}{\ensuremath{\texttt{select} }\xspace}
\newcommand{\ratio}{\ensuremath{\rho }\xspace}

\newcommand{\insertion}{\ensuremath{\texttt{INSERT} }\xspace}
\newcommand{\update}{\ensuremath{\texttt{UPDATE} }\xspace}
\newcommand{\delete}{\ensuremath{\texttt{DELETE} }\xspace}

\newcommand{\sysfull}{AlphaClean\xspace}
\newcommand{\sys}{AlphaClean\xspace}
\newcommand{\sysnospace}{AlphaClean}

\newcommand{\tbl}[1]{\textsf{#1}\xspace}
\newcommand{\field}[1]{\textsf{#1}\xspace}
\newcommand{\cost}{\textrm{cost}\xspace}
\newcommand{\ans}{\textsf{ans}\xspace}
\newcommand{\dans}{\Delta\textsf{ans}\xspace}
\newcommand{\cqp}{correction query processing\xspace}
\newcommand{\Cqp}{Correction query processing\xspace}

\newcommand{\reminder}[1]{{{\textcolor{magenta}{\{\{\bf #1\}\}}}\xspace}}
\newcommand{\ewu}[1]{{{\textcolor{blue}{\{\{\bf ewu:\} #1\}}}\xspace}}
\newcommand{\mps}[1]{{{\textcolor{red}{\{\{\bf meelap:\} #1\}}}\xspace}}
\newcommand{\stitle}[1]{\smallskip\noindent\textbf{#1: }}
\newcommand{\ititle}[1]{\smallskip\noindent\textit{#1: }}
\newcommand{\btitle}[1]{\smallskip\noindent\textbf{#1}}

\definecolor{light-gray}{gray}{0.95}
\definecolor{mid-gray}{gray}{0.85}
\definecolor{green}{RGB}{0,176,80}
\definecolor{darkred}{rgb}{0.7,0.25,0.25}
\definecolor{darkgreen}{rgb}{0.15,0.55,0.15}
\definecolor{darkblue}{rgb}{0.1,0.1,0.5}
\definecolor{orange}{RGB}{237,125,49}
\definecolor{blue}{RGB}{68,114,196}
\definecolor{pop}{RGB}{0,21,245}

\newcommand{\white}[1]{{\textcolor{white}{#1}\xspace}}
\newcommand{\blue}[1]{{\textcolor{blue}{{\bf #1}}\xspace}}
\newcommand{\orange}[1]{{\textcolor{orange}{{\bf #1}}\xspace}}
\newcommand{\pop}[1]{{\textcolor{pop}{{\textit{\textbf{#1}}}}\xspace}}
\newcommand{\red}[1]{\textcolor{red}{#1}}
\newcommand{\green}[1]{\textcolor{green}{#1}}
\newcommand{\gray}[1]{\textcolor{light-gray}{#1}}

\newcommand{\specialcell}[2][c]{%
  \begin{tabular}[#1]{@{}c@{}}#2\end{tabular}}

\def\ojoin{\setbox0=\hbox{$\bowtie$}%
  \rule[-.02ex]{.25em}{.4pt}\llap{\rule[\ht0]{.25em}{.4pt}}}
\def\leftouterjoin{\mathbin{\ojoin\mkern-5.8mu\bowtie}}
\def\rightouterjoin{\mathbin{\bowtie\mkern-5.8mu\ojoin}}
\def\fullouterjoin{\mathbin{\ojoin\mkern-5.8mu\bowtie\mkern-5.8mu\ojoin}}

\title{DeepLens: Towards a Visual Data Management System}

%\numberofauthors{1}
\author{ Sanjay Krishnan, Adam Dziedzic, Aaron J. Elmore \\
\affaddr{ University of Chicago} \\
\affaddr{ \{skr, ady, aelmore\}@uchicago.edu}\\
}

\fontsize{9.5pt}{11pt}
\selectfont

\maketitle

\begin{abstract}
Advances in deep learning have greatly widened the scope of automatic computer vision algorithms and  enable users to ask questions directly about the content in images and video.
This paper explores the necessary steps towards a future Visual Data Management System (VDMS), where the predictions of such deep learning models are stored, managed, queried, and indexed.
We propose a query and data model that disentangles the neural network models used, the query workload, and the data source semantics from the query processing layer.
Our system, \textsf{DeepLens}, is based on dataflow query processing systems and this research prototype presents initial experiments to elicit important open research questions in visual analytics systems.
One of our main conclusions is that any future ``declarative'' VDMS will have to revisit query optimization and automated physical design from a unified perspective of performance and accuracy tradeoffs.
Physical design and query optimization choices can not only change performance by orders of magnitude, they can potentially affect the accuracy of results.
\end{abstract}

%\pagenumbering{gobble}

\section{Introduction}\label{intro}\sloppy
Recent advances in deep learning have enabled new forms of analysis on images and videos~\cite{lecun2015deep}. 
A typical analytics task is to find all images in a corpus that contain a particular object (e.g., detecting a red car in CCTV footage) using neural network-based object detection models.  
Answering such queries at scale is a formidable computer systems challenge with significant recent interest from both industry and academia~\cite{lu2018accelerating, lu2016optasia, kang2017noscope, anderson2018predicate, kang2018blazeit, chetlur2014cudnn, fengeva, zhang2018ffs, jiang2018mainstream, jiang2018chameleon}. 

Recent work on visual analytics is clearly inspired by ideas from relational database systems~\cite{lu2016optasia,kang2018blazeit,kang2019blazeit}: there is a high-level query language and an optimized execution layer that processes the queries.
\emph{But, are these new systems truly declarative like an RDMBS?}
We argue that existing systems neglect some of the critical challenges in the space; giving an illusion of a declarative interface for simple query processing tasks.
This paper explores the execution tradeoffs in visual analytics and illustrates a complex interplay storage, latency, and accuracy considerations.
Abstracting these tradeoffs away from an end-user will be the primary technical challenge in a future Visual Data Mangement System (VDMS).

An obvious first omission from existing proposals is a principled study of input data formats.
Today's systems interface with video data \emph{as independent frame-ordered sequences of raw images}. While this format is convenient for implementation, it may not be the optimal physical layout for all use cases.
Video data is often stored and transferred in an encoded formats, such as H.264, which can significantly reduce the storage size of high-resolution video (sometimes by multiple orders of magnitude).
On the other hand, due to the sequential nature of most video encoding schemes (where decoding requires scanning preceding frames), encoding precludes random access to particular frame numbers or times in the video--affecting possible execution paths to answer a given query. 
Furthermore, most encoding schemes are lossy, which leads to downstream accuracy considerations after decoding.
One could also imagine hybrid physical layouts that segment the video into short encoded clips and then bucket them by time.
The complexity of this space means that any VDMS has to allow for a multitude of encoding and physical layout schemes, and decouple these layouts from the programming model.
 
Beyond just physical layout, answering complex queries in such systems is challenging.
Consider a variant of the task before: given two videos find a particular object that appears in both videos. 
For example, we might be interested in determining if the same car appeared in two different CCTV feeds.
To answer this query, one has to first find potential target objects in both videos and then match them against each other. 
Even systems that support \emph{join queries} across videos~\cite{lu2016optasia}, have difficulty answering questions when the join predicate considers pixel data.
To be able to answer such a query efficiently, there are implementation decisions about indexing (can a multidimensional index be used for faster matching), optimization (if we do index, which video to scan and what type of an index to use), and compression (can the matching be performed on low-dimensional features instead of raw frames). 
Indexing and query optimization are significant missing components in current proposals, which are necessary to scale such systems to more complex analytics tasks.

To go beyond these limitations, this paper explores the missing pieces towards a full-featured VDMS.
For the programming abstraction, we start with a analog to the relational model.
All visual corpora can be described as an unordered collection of subimages (called patches) and an associated key-value dictionary storing information about them (e.g., neural net classifications, their provenance).
These patches are abstract data types (stored in raw pixels or pre-compressed to features) that can be generated from object detection models that crop identified objects, classical visual segmentation methods that crop based on color, or even whole images.
The query model, which allows for Selects and Joins over patch collections, has set semantics as in SQL and makes no structural assumptions about the data source, e.g., any timing data in video or geometric data from camera correspondences are stored as additional attributes.
In other words, every operator defines closed-algebra: collection of patches in and collection of patches out.
Optimizations for specific workloads are introduced through physical design: data layout, and both single-attribute and multi-dimensional indexing on patch collections.

\begin{figure}[t]
%\vspace{-5pt}
\centering
 \includegraphics[width=\columnwidth]{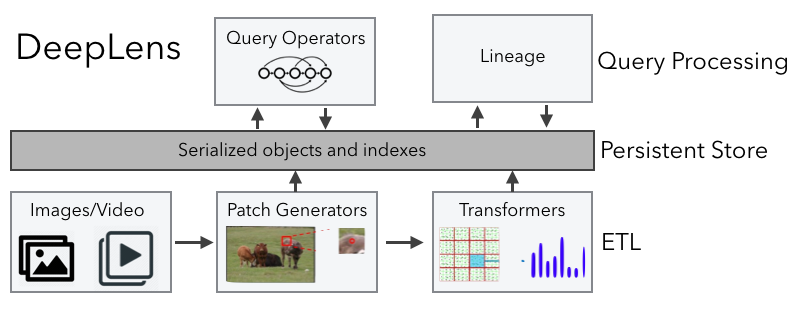}
 \caption{DeepLens has a dataflow-like architecture for processing visual analytics queries. All analysis is cast as relational queries on relations of image patches. Intermediate results can be materialized and indexed.  \label{teaser} }
\end{figure}

This logical layer of indirection disentangles the process that generates the patches (decoding, neural network inference, segmentation) with the downstream query processing of the collection.
The query operators are agnostic to whether those patches represent the output of a neural network, black-box geometric computer vision pipeline, or even the output from crowdsourcing. 

We built a research prototype system, \textsf{DeepLens},
that implements this model. 
To illustrate the benefits of physical and logical separation, we design a benchmark of six analytics tasks and vary what is indexed, how the operators are implemented, the underlying algorithms, and the properties of the underlying hardware.
A substantial number of queries of interest require matching predictions across multiple videos or matching outputs across multiple prediction pipelines, and our central insights is that \emph{indexing} and \emph{lineage} are critical missing pieces from the current discourse on visual analytics.
 
\subsection*{Key Insights}

\vspace{0.5em} \noindent \textbf{Encoding: } Compressing video data with standard codecs can save up to 50x in storage, but requires additional computation and scan costs to process. Segmenting video into clips and then compressing those clips provides a flexible way to optimize this tradeoff between storage and computation.

\vspace{0.5em}  \noindent \textbf{Indexing: } We discovered that index usage is crucial and improves query performance by up-to 612x. However, particularly for the geometric indices, the size and dimensionality of the data plays an essential role in how beneficial an index will be.

\vspace{0.5em}  \noindent \textbf{Lineage: } We found that many tasks require relating processed results back to the base data.
Maintaining and indexing tuple-level lineage led to a 60x improvement in one benchmark query.

\vspace{0.5em}  \noindent \textbf{CPU vs. GPU: } Cost-models that accurately balance CPU and GPU utilization for query optimization will be a significant challenge. Some queries benefited by nearly two orders of magnitude, while others are slower when processing tasks are offloaded to a GPU.

\vspace{0.5em} \noindent \textbf{Managing Uncertainty: } Unlike in relational databases, queries in a VDMS are approximate by nature. Traditional optimizations, like filter pushdown, may additionally create different accuracy profiles.  

\vspace{0.5em}
One of our main conclusions is that any future ``declarative'' VDMS will have to \emph{revisit query optimization and automated physical design from a unified perspective of performance and accuracy tradeoffs.}

\section{Background and API}
As a direct consequence of the rapid progress in computer vision, visual analytics workloads are becoming increasingly complex with requirements. 
One can easily make brittle assumptions about ordering of images or the structure of any associated metadata.
Or, one might use ad-hoc data structures that will not scale past memory limits.
This echoes the challenges of the pre-relational days of database research before the idea of data independence where the query processing, the data, and the storage format are all intertwined.
We want a unified model that covers much of visual analytics that can allow us to leverage logical-physical separation principles that the database community has pioneered.

\subsection{The ``Narrow Waist'' of Visual Analytics}
DeepLens is based on a logical model where data are unordered collections of image patches (featurized sub-images) with their associated derivation information and metadata.
The user writes programs against such a logical model and a query optimizer selects an appropriate execution path.
The patches and metadata are typed to allow the system to validate logical equivalence.
This architecture abstracts lower level concerns about the physical layout, while simultaneously ensuring that the system can reason about the application logic.
Physical concerns include video encoding, temporal segmentation, and cross-video indexing.
Application-level concerns include data lineage, whether a component (e.g., an object detector) can be replaced with another one, and latency-accuracy tradeoffs.
We envision that such a narrow waist design is crucial as we scale visual analytics to harder problems, such as in robotics where it is increasingly common to integrate predictions from multiple neural networks~\cite{hodson2018robots}.
\vspace{-0.25em}
\subsection{Iterator API for Patches}
After surveying a number of use cases from robotics to traffic camera analysis, we arrived at a simple query model, where visual analytics queries are relational queries over collections of subimages (called patches) and associated metadata about how those subimages were generated.
The query processing engine is agnostic to how those patches are generated.
They can be whole images, smaller tiled subimages, or even subimages extracted by an object detection neural networks.
In pseudocode, a \texttt{Patch} object contains a pointer to the image that generated it, the data contained in the patch, and a key-value metadata dictionary:
\begin{lstlisting}
Patch(ImgRef, Data, MetaData)
\end{lstlisting}
We are purposefully vague about the structure of \texttt{Data} and \texttt{MetaData}. We only assume that \texttt{Data} is an n-dimensional dense vector (can represent pixel content or image features) and \texttt{MetaData} has a key-value dictionary of attributes of this data.
Inspired by dataflow systems, operators in the system implement iterators over tuples of \texttt{Patch} objects:
\begin{lstlisting}
Operator(Iterator<Tuple<Patch>> in, 
         Iterator<Tuple<Patch>> out)
\end{lstlisting}
Lineage is maintained as every operator is required to update the \texttt{ImgRef} attribute to retain a lineage chain back to the original image.

\subsubsection{Example 1.}
Let us first consider an example studied in existing work, such as~\cite{lu2016optasia,lu2018accelerating,kang2018blazeit}.
Consider a CCTV feed of a parking lot that collects and stores video.
We want to evaluate the parking lot's utilization so we want to count the number of cars in each frame of the video.
We first run all of the frames through the Single-Shot Detector (SSD) object detection network~\cite{liu2016ssd}, which returns bounding boxes and labels for common objects detected in the image.
Each of these bounding boxes defines a patch:
\begin{lstlisting}
SSDPatch(Frame, Bbox, 
         {'label': L, 'frameno': F })
\end{lstlisting}
Over these SSD patch objects, the query of interest is well expressed in relational algebra over the metadata dictionary (a filter over labels and an aggregation over frame numbers).

\subsubsection{Example 2}
Now, let us consider a marginally more difficult example.
Suppose we are given two videos from two different cameras, we want to find all cars that appear in both videos.
The first step is exactly the same as the previous example, we break the frames up into \texttt{SSDPatch} objects.
Over the two sets of \texttt{SSDPatch}, we need to compare all of the bounding boxes and return those that are sufficiently similar in terms of image content.
What is different about this query is that it leverages both the metadata and the pixel data in the bounding boxes.

Interestingly, there are a number of unresolved questions about how to actually process this query efficiently.
Naively, one could compare all pairs of bounding boxes and then return those of sufficient image similarity.
Most image matching algorithms use lower dimensional features to match, so another option is to pre-compute the relevant features and build a multidimensional index over one of the sets of \texttt{SSDPatch} objects, e.g., a KD-Tree over a set of color histograms.

\subsection{Related Work}
The new era of ``deep'' multimedia systems are not actually multimedia database systems in the traditional sense~\cite{faloutsos2012searching}. In current systems, the queries issued to pixel data are very restricted: either binary neural network predicates~\cite{kang2017noscope, zhang2018ffs, anderson2018physical, jiang2018chameleon} or highly structured~\cite{lu2016optasia,lu2018accelerating}.
Our work is inspired by prior work on multimedia databases, which have long acknowledged the importance of indexing and query processing strategies~\cite{yoshitaka1999survey, faloutsos2012searching}.
\textsf{DeepLens} revisits the idea of a multimedia database in the era of deep learning, where the content--both structured and unstructured--is populated by the outputs of a neural network inference pipeline.
Recent work can be best summarized as filter optimization~\cite{kang2017noscope, zhang2018ffs, anderson2018physical, jiang2018chameleon}; how to evaluate a neural network predicate as quickly as possible while satisfying an accuracy constraint.
The Optasia system~\cite{lu2016optasia} goes beyond simple predicates and does consider joins; however, as noted earlier the joins do not efficiently handle predicates over the image data (not just the metadata).
We project into the future that the community will soon move past filters and also consider joins and more complex query operators.
Addressing the more complex visual analytics queries will require leveraging ideas from classical work on indexing strategies~\cite{faloutsos2012searching}.
However, problems in this new setting are higher dimensional and have to manage uncertainty.
Another important design decision was to natively manage image transformation and provenance code in our system.
We chose this architecture since recent work manipulates neural network structure, e.g., by cascading models to improve accuracy or performance~\cite{kang2017noscope, anderson2018physical, jiang2018mainstream}.
Another recent paper that we would like to highlight is a data management system for Augmented Reality~\cite{haynes2018lightdb}, and many of the ideas relating to multidimensional indexing will be very relevant.
This paper evaluates all of these design trade-offs and proposes an initial architecture for a modern VDMS.
One of our main conclusions is that any successful VDMS system will have to feature a sophisticated and robust query optimizer and some level of automated physical design. 
We envision that new results in learning-based query optimization~\cite{kaftan2018cuttlefish,krishnan2018deeprljoins} and ideas inspired by the new automated tools for physical design such as~\cite{sharma2018case,pavlo2017self} will be crucial.

\section{Storage Layer}
Our architecture is designed as a middle ground between the pure stateless dataflow approach and the RDBMS approach.
\textsf{DeepLens} provides dataflow operators to shape, transform, and manipulate image corpora at scale, but also supports materialization and indexing of intermediate results.
Figure \ref{teaser} illustrates the basic architecture with three main layers: (1) Persistent Storage, (2) ETL,  and (3) Query Processing.
We first describe the storage layer in \textsf{DeepLens}.

\subsection{Raw Data}
In our initial prototype, we only consider videos at rest and not streaming videos. Our loading API is as follows:
\begin{lstlisting}
Load(filename, filter=True)
\end{lstlisting}
Videos are loaded into the system returning an iterator that returns a patch collection where each patch is a full video frame: 3-D dense arrays representing width and height of images, as well as  corresponding RGB channels.
The loader can take a filter as an optional argument and it only returns those frames that satisfy the filter condition.
The loader abstracts the encoding scheme of the underlying video from the user. \textsf{DeepLens} supports the following storage formats:

\vspace{0.5em} \noindent \textbf{Frame File: } In the most basic format, we treat each frame of a video as a single record. This frame can either be stored in its raw pixel values or in some featurized format. The \emph{Frame File} is implemented with BerkeleyDB~\footnote{Implemented with bsddb3 (Python binding for BerkeleyDB)}. The image and feature data is serialized in a binary format before insertion.  By default, they are stored in a sorted file by frame number for videos and wall clock time. This is because many queries of interest examine particular time segments of videos. The sorted file allows for quick retrieval of temporal predicates. The advantage of the Frame File is a temporal filter push down; the disadvantage is that it can require significantly more storage for video. For example, the H.264 codec often reduces the storage footprint of video by 20+ times. Storing video as a sequence of image frames precludes the use of sequential encoding techniques. 

\vspace{0.5em} \noindent \textbf{Encoded File: } Therefore, we also support storing video in common encoded formats (e.g., OGG, MPEG4). The tradeoff is that encoding precludes pushing down temporal predicates since many encoding formats require a sequential decoding procedure. Furthermore, the cost of decoding the video must also be factored in.

\vspace{0.5em} \noindent \textbf{Segmented File: } As a hybrid between the Frame File and the Encoded File, we have the Segmented File. This storage format segments the video into short clips and stores the encoded clips in BerkeleyDB. We can benefit from coarse-grained temporal filter push down, while having some benefits of encoding. 

\subsection{Derived Data}
Any of the intermediate results in \textsf{DeepLens} can be materialized as well. 
We also support the construction of indexes on the materialized data.
The challenge is that every data type requires a specialized index structure.
Over string valued or discrete metadata, the index choices are straight-forward. 
We support hash tables and B+ Trees over any key (both of which are implemented with BerkeleyDB).

For the multidimensional data (patch segmentation parameters or image features), indexing is a little more nuanced and dependent on the workload.
As a concrete example, let us consider patches that are parametrized by ``bounding boxes'' (x1,y1,x2,y2).
Even though this data is multidimensional, we might effectively be interested in single dimensional queries.
For example, find bounding boxes left of a certain point in the image.
To support such queries, we found that it was far more efficient to use a B+ Tree rather than an R-Tree multidimensional index.
On the other hand, if we are interested in containment and intersection queries, then we need a true multidimensional index.
We provide an interface to a disk-based R-Tree implemented with \texttt{libspatialindex}\footnote{https://libspatialindex.github.io/}.

However, we found that existing R-Tree implementations are optimized for geospatial problems in 2D.
They could not be efficiently modified for higher dimensional data, for instance, for image matching queries, where we compare features of two images and threshold the similarity.
For this class of queries and data, a data structure called a Ball-Tree was the most effective at answering Euclidean threshold queries in high-dimensional spaces~\cite{kumar2008good}.

In summary, \textsf{DeepLens} supports a large number of single-dimensional (Hash, B+ Tree, Sorted Files) and multi-dimensional indexes (R-Tree and Ball-Tree). The physical design problems are interesting as even the same attributes might be indexed in different ways depending on the workload.

\subsection*{Future Work: Storage Advisor}
In this paper, we prototyped all of the different storage options, and manually selected the appropriate ones per use case. In the future, we would like to have a storage advisor that can analyze a workload or an SLO and return an optimized storage scheme. We see an analog with the work that has been done in the database community on materialization and storage format tuning~\cite{agrawal2006automatic}.
\section{Visual ETL}
Once loaded, the image dataflow is passed into \textbf{patch generators} that turns the image into a set of patches. Associated with each patch is a key-value dictionary storing metadata about how the patch was generated and any property of the patch. The patches are then fed into a composition of \textbf{transformers} that featurize, compress, or otherwise store detected properties of the patches into the dictionary. Over tuples of transformed patches, users can build a directed computation graph with \textbf{operators} (e.g., select, join).
Tuple-level lineage is automatically maintained by the system allowing any downstream patch to be associated with its base data.
Any of the intermediate results can be \textbf{materialized}. 

\vspace{1.5em}
\subsection{Visual ETL}
\label{subsection:visualETL}
 In a sense, image data is unstructured data. Semantics from image data have to be first extracted with computer vision algorithms before structured queries can be executed. 
The ETL layer defines the generation of \texttt{Patch} objects and their manipulation.

\vspace{0.5em}
\noindent \textbf{Patch Generators: } We provide a library of \texttt{Patch Generators}. 
These generators take as input an iterator over raw images and return an iterator over \texttt{Patch} objects.  
Our experiments consider three instantiations of these generators: object detection (ones that segment objects out of the image return segmentation masks), optical character recognition (one that identifies text in an image and returns a segmentation as well as the recognized strings) and whole-image patches (returns the whole image).

\vspace{0.5em}
\noindent \textbf{Transformers: } The main content of a patch is still raw pixel data, which is often not very useful on its own.
To be able to compare or manipulate the patches, we need a featurized representation.
This leads to the next module of the ETL layer that defines Transformers. 
A transformer takes as input an iterator over \texttt{Patch} objects and returns an iterator over transformed \texttt{Patch} objects.  
In our experiments, we consider two transformers: color histogram features for image matching and a depth prediction neural network which predicts the 3D geometry of a patch. 

\vspace{0.5em}
\noindent \textbf{Materialize: } \textsf{DeepLens} allows any stage of this pipeline to be materialized and persisted to disk or memory.

\subsection{Validation}
The entire API in \textsf{DeepLens} is typed, which allows us to validate pipelines. The computer vision community is rapidly changing, and we foresee new visual processing primitives will constantly be added to any VDMS.
Beyond the standard \texttt{int}, \texttt{float}, \texttt{string} types, 
our type system maintains the resolution and dimensions of each patch (almost all neural networks used today require fixed input resolutions). We also include the domains of any discrete metadata created when available. For example, object detection networks have a closed-world of labels that they could predict. We include this information in the type system. Any downstream operator (e.g., filter) that consumes those labels can be validated to see if that label is plausibly produced by the pipeline. 

\subsection*{Future Work: Pipeline Synthesis}
Long term, we would like a system that could declaratively synthesize a pipeline given a library of generators and transformers and latency/accuracy constraints. Concretely, suppose we had a library of general purpose pre-trained object detection models and some special case programmed models (e.g., a car detector). 
We envision a system that scores each model with a precision/recall profile for a desired dataset, and can choose the model that is most appropriate for a query.
This can only be done if there is an appropriate type system that understands what labels can be predicted and when detectors are interchangeable.

\vspace{1.5em}
\section{Query Processing}
Our query processing engine is designed like a dataflow query processing system~\cite{graefe1994volcano}.
In our initial prototype, we implement \texttt{Select} and \texttt{Join} operators. The design of the \texttt{Join} operators are the most interesting so we highlight it here.  

\vspace{0.5em}
\noindent \textbf{Nested Loop Join: } If no indexes are available, the most generic operator is a nested loop join operator. This join operator can execute arbitrary $\theta$-joins on the data.
It compares all pairs of patches from two collections and returns those that satisfy a predicate.

\vspace{0.5em}
\noindent \textbf{Index Joins: } If a multi-dimensional or single dimensional index is available, we can use that index to enable equality joins, range joins, or similarity joins.

\vspace{0.5em}
\noindent \textbf{On-The-Fly Index Similarity Join: } We found that for image matching queries where one of the relations was relatively small, the index could be constructed on-the-fly.
We load the smaller relation into an in-memory Ball-Tree. Then, probe using the other collection of patches.

\subsection{Lineage}
Many visual analytics tasks of interest relate processed results back to the base data.
For example, we might process a single set of images in two different ways, e.g., segmenting the image with an object detector and using a depth prediction model to determine relative distance between pixels.
To relate these results, we have to run a \emph{backtracing} query (select all raw images that contributed to a patch).
This is similar to queries studied in recent lineage systems~\cite{psallidas2018smoke}. 

\textsf{DeepLens} natively tracks tuple-level lineage.
Every \texttt{Patch} object maintains a descriptor how it was generated from either a raw image or another patch.
Its relationship to the base data is maintained as a sequence of pointers.
This information is stored as attributes in the metadata key-value dictionary so indexes and queries can be natively supported on them.

\subsection*{Future Work: Visual Query Optimizer}
We believe that the key to a true VDMS will be a query optimizer than can manage and abstract all of the complexity of the physical operations.
Classical database operators are mostly I/O bound, but on the other hand, many VDMS operators are compute-bound.
The caveat is that device placement (CPU/GPU) matters for these operators.
Cost-models that accurately balance CPU and GPU utilization for query optimization will be a significant challenge. In our experiments, some queries benefited by nearly two orders of magnitude by GPU optimizations, while others are actually slower when the processing tasks are offloaded to a GPU. We hope to explore new strategies for joint multi-query optimization and optimal physical design. Doing so might well require significant machine learning to navigate a noisy and analytically complex cost model. 
Image queries are always approximate, and managing uncertainty and quantifying the accuracy effects of a certain plan will be a serious challenge.

\vspace{1.5em}
\section{Benchmarks}
We briefly discuss how we designed the benchmark datasets and corresponding workload.

\subsection{Datasets}
From publicly available sources, we constructed datasets to simulate envisioned use cases for \textsf{DeepLens}. An important consideration was images of varying format and size.  We decided that it was important to evaluate a diversity of tasks and the robustness of the injestion pipeline to different formats, sizes, and content. 

\vspace{0.5em} \noindent \textbf{PC.} This dataset is designed to simulate a dataset of images found on a personal computer. It consists of 779 photographs, screenshots, and document scans. 

\vspace{0.5em} \noindent \textbf{TrafficCam.} This dataset consists of 24 mins and 30 secs of high-definition (1080p) traffic camera video (35280 frames).

\vspace{0.5em} \noindent \textbf{Football.} This dataset consists of 15 low-definition (720p) videos of American football clips of the same team ranging from 30 secs to 1 mins (15244 total images). 

\subsection{Queries}
We propose a benchmark workload of 6 queries on these datasets to evaluate \textsf{DeepLens}. These queries are inspired by problems considered in prior work and consist of tasks that involve querying pixel data, the results of neural network predictions, relating results back to base data, and combinations of these tasks.

\vspace{0.5em} \noindent \textbf{q1.} \emph{Find all near-duplicates in the PC dataset}. This query is inspired by classical multimedia information retrieval problems such as reverse image search (find the closest image to a query image). 

\vspace{0.5em} \noindent \textbf{q2.} \emph{Count all of the frames with at least one vehicle present in the TrafficCam dataset}. This query is inspired by recent work that uses neural networks to analyze traffic and movement patterns~\cite{kang2017noscope}. This is a simple query that takes the output of a neural network that identifies objects in the frame and simply queries the output.

\vspace{0.5em} \noindent \textbf{q3.} \emph{Track one player's trajectory in every play in the Football dataset}. Given segmentation output that identifies a player in frame and OCR output that identifies a number if one is visible, we have to relate that sequence of bounding boxes back to the original image.

\vspace{0.5em} \noindent \textbf{q4.} \emph{Count all distinct pedestrians in the TrafficCam dataset}. This query is a variant of q2. The distinct qualifier makes this query significantly more challenging as it requires deduplicating candidate pedestrians detected in the video.

\vspace{0.5em} \noindent \textbf{q5.} \emph{Lookup the presence of a string in the PC dataset}. Apply OCR to all of the images, and store a collection of strings discovered. We run a query to identify the first image with a target string.

\vspace{0.5em} \noindent \textbf{q6.} \emph{Find all tuples of pedestrians (p1,p2) where p1 is behind p2 in the TrafficCam dataset}. This query is inspired from applications in robotics and navigation, where one has to estimate how far a given object is located from a camera. This problem, called depth prediction, has recently been a subject of research interest in computer vision~\cite{depthPredictModel}. We leverage the published code\footnote{https://github.com/iro-cp/FCRN-DepthPrediction} and the pre-trained parameters to annotate all detected pedestrians in the TrafficCam dataset with depth predictions and find such pairs.

\section{Experiments}
Recent work on video and image analytics systems has largely focused on optimizing neural network inference.
For queries that are simply filters or aggregates, neural network inference clearly dominates the processing cost.
In particular, we use our benchmark to understand the execution profiles of more complex queries where the subsequent processing time before or after neural network inference might dominate.

\begin{figure}[t]
% \vspace{-5pt}
\centering
 \includegraphics[width=0.75\columnwidth]{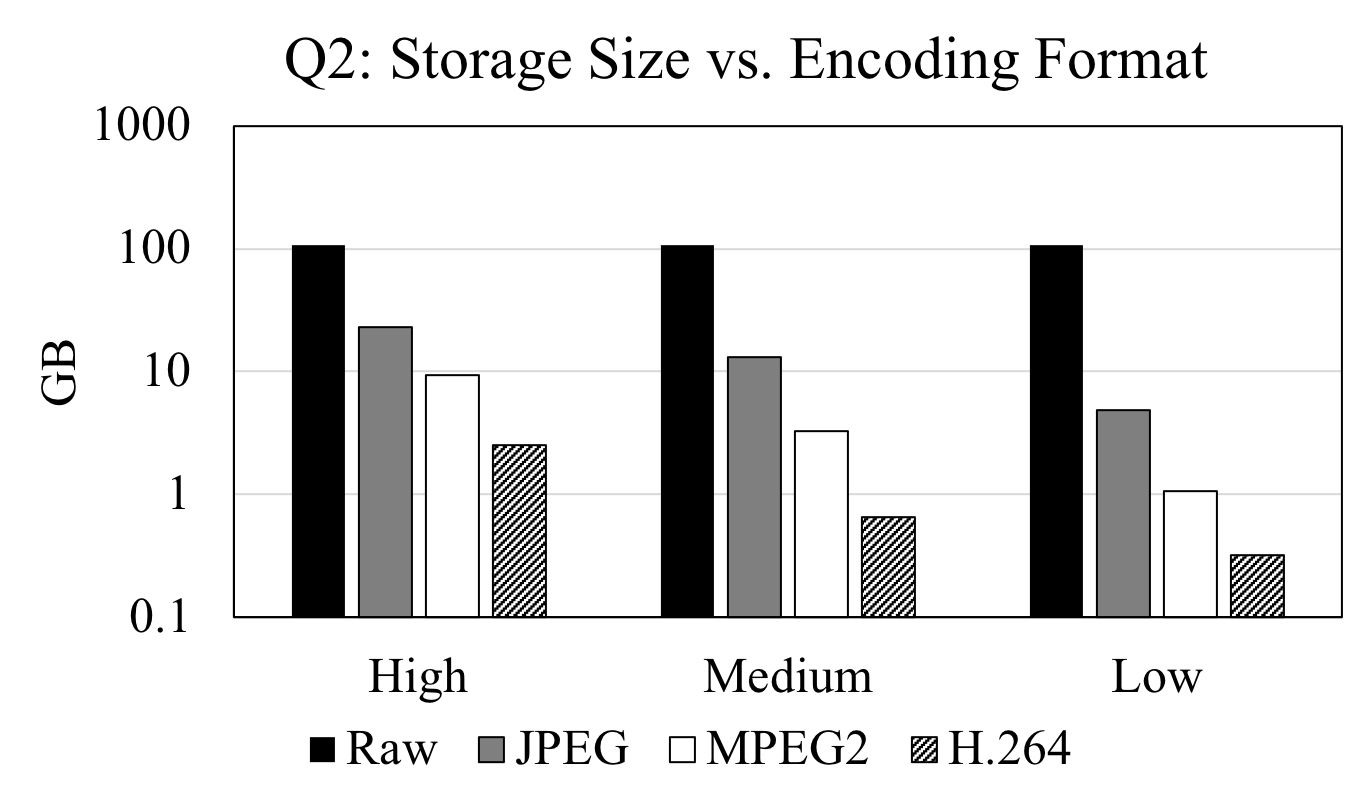}
 \includegraphics[width=0.75\columnwidth]{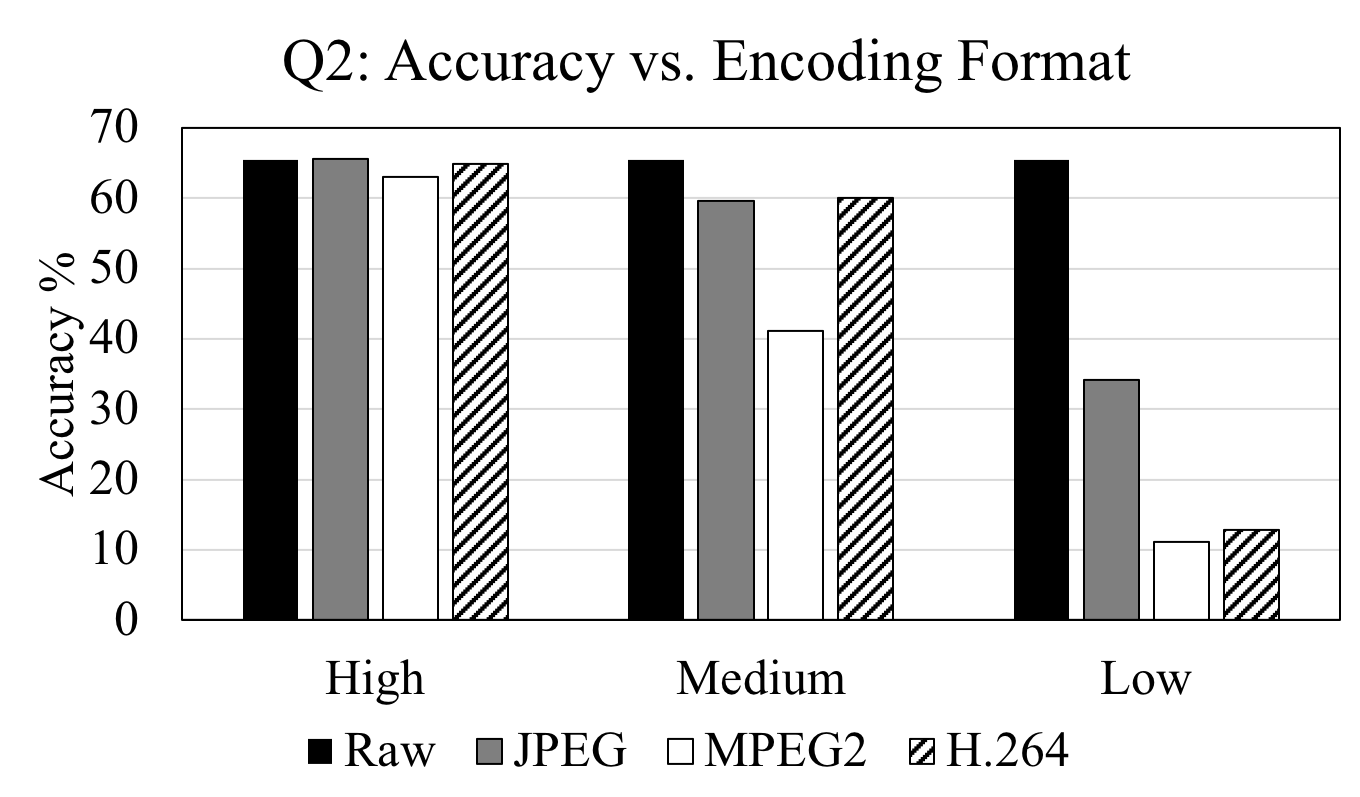}
 \caption{Encoding a video with a sequential codec can reduce storage costs by over 50x without loss of accuracy. \label{encoding} }
\end{figure}

\subsection{Data Encoding}
In the first experiment, we illustrate some of the interesting challenges with data encoding. 
We compare different encoding formats for the video in Q2 in terms of storage cost as well as the overall accuracy of the pipeline (Figure \ref{encoding}).
We estimated the accuracy by manual annotation of the video.
We compare three levels of lossy encoding: High, Medium, Low.

\begin{figure}[t]
% \vspace{-5pt}
\centering
 \includegraphics[width=0.80\columnwidth]{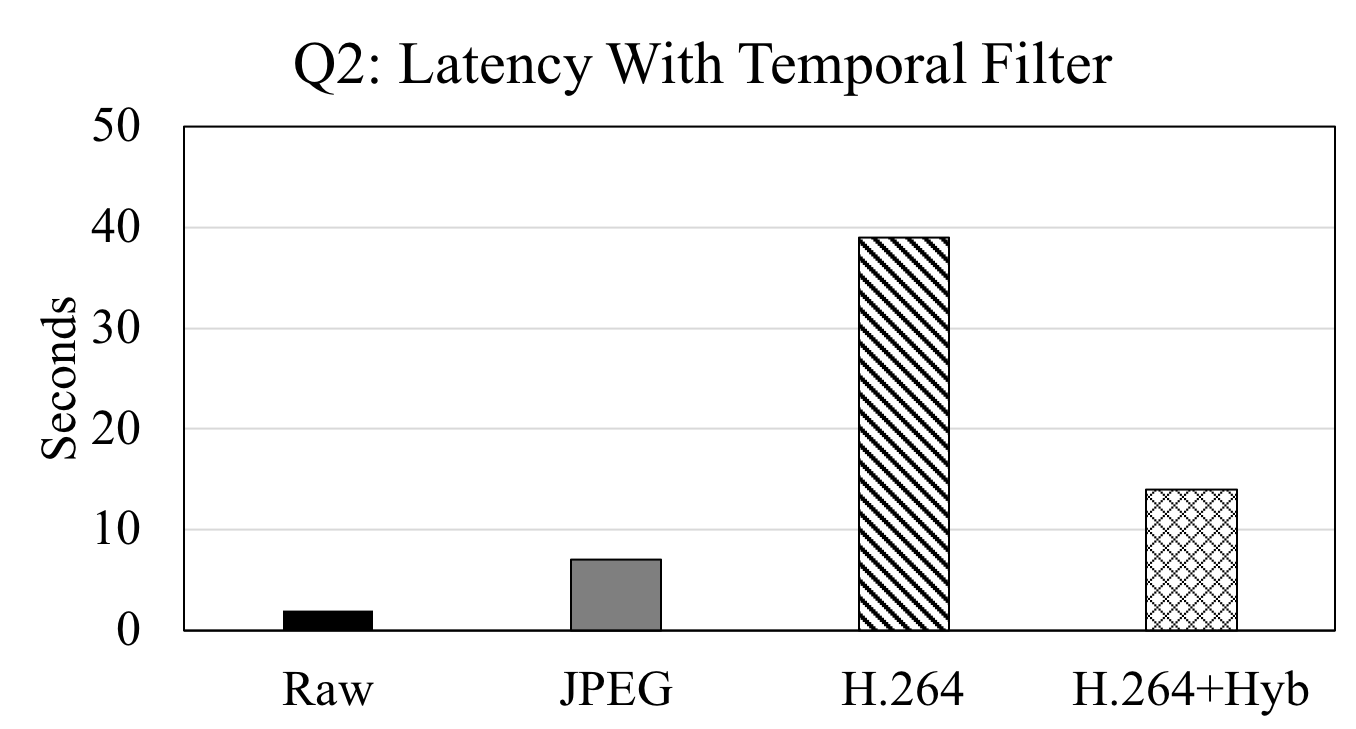}
 \caption{Hybrid storage formats can support coarse-grained filter push down as well as take advantage of sequential compression. \label{encoding2} }
\end{figure}

The RAW encoding (where every frame is an image) rests at about 107 GB on disk.
In contrast, a state-of-the-art H.264 encoding\footnote{Implemented with https://github.com/cisco/openh264} only takes up 2.5 GB on disk.
These encodings are all lossy, but there is negligible impact on downstream accuracy for the "high quality" encodings.
For larger compression ratios, we do see a degradation in downstream performance.

Of course, there is a latency consideration as well.
We evaluate a subset of the input formats in terms of their end-to-end latency (including decoding) in Figure \ref{encoding2}.
We add a temporal filter to Q2 (selecting a small subset of frames to analyze).
The H.264 encoding cannot support a true filter push down as the codec algorithm is sequential and requires a scan of the preceding video until that point.
On the other hand, the JPEG and RAW formats can trivially support the push down optimization.
A hybrid scheme that breaks the video into short clips and then encodes them, can allow for coarse-grained filter push down. 
We manually tuned this granularity for best performance.

These encoding experiments illustrate the complexity of visual analytics physical layout problems.
VDMSes must support a variety of these formats and structures to be effective.
We hope this is also an opportunity to re-visit automatic storage tuning and declarative interfaces for storage configuration.

\begin{figure}[t]
% \vspace{-5pt}
\centering
 \includegraphics[width=0.8\columnwidth]{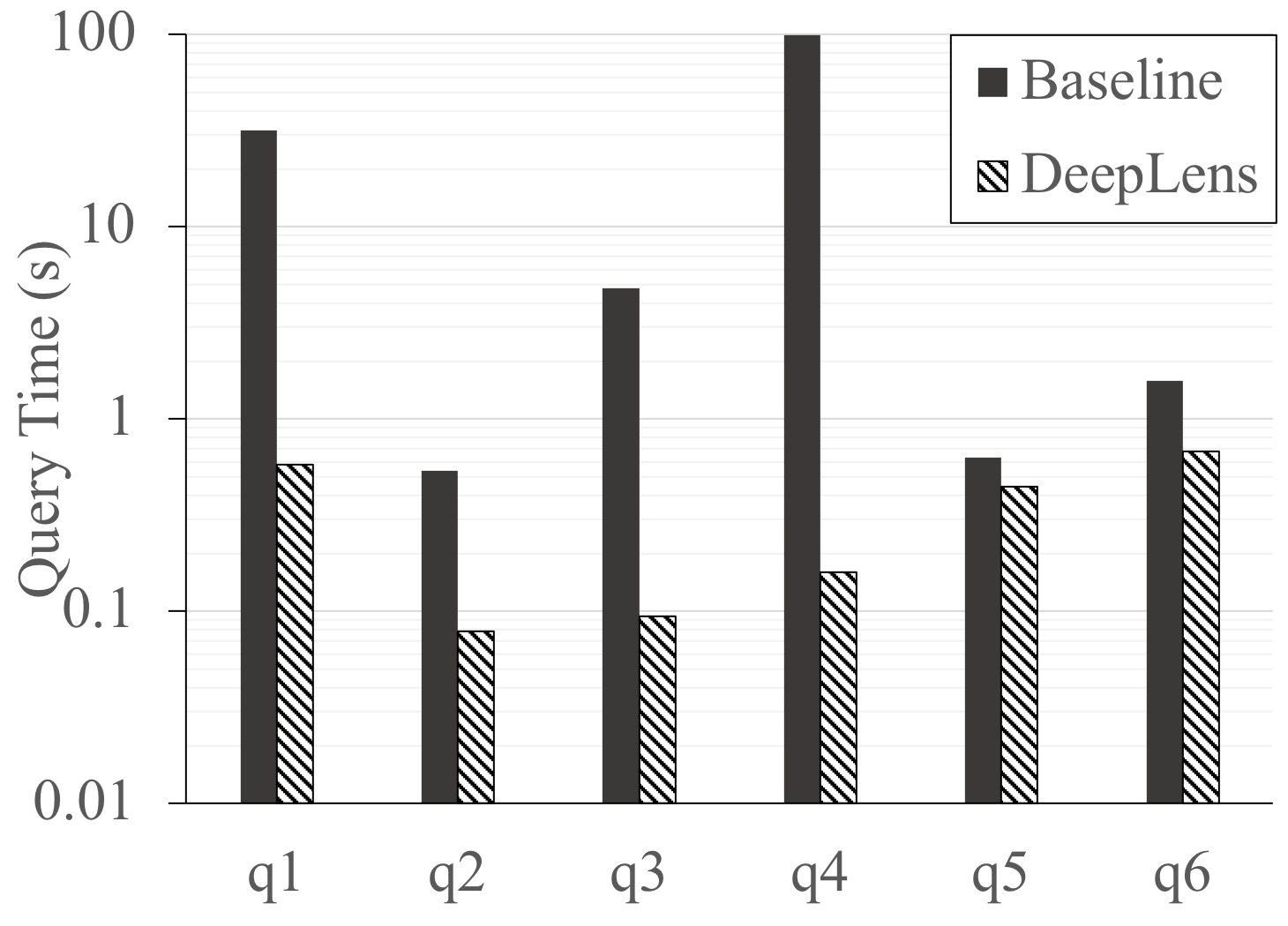}
 \caption{\textsf{DeepLens} significantly speeds up ``query time'' by using indexes. The queries that match multidimensional features can be sped up by up-to 600x.  \label{query} }
\end{figure}

\subsection{The Power of Indexes}
Next we study indexing and materializing intermediate results.
In these experiments we separate ``Query time'' (processing patches) from ``ETL time'' (generating patches).
Many ETL processes can be reused for multiple queries, and those costs can be amortized. 

Figure \ref{query} plots the query times with and without indexing.
Our baseline is the same query processing engine with no indexes.
We compare this baseline to a hand-tuned version where we manually select the best physical design for a query.

The queries that benefit the most from the indexes are ones that require image matching, and are up-to 612x faster for q4 and 59x faster for q1. Note that these matching costs are incurred after neural network inference, so techniques such as~\cite{kang2017noscope, anderson2018predicate, kang2018blazeit} would have no effect on this step.
Queries that have to relate results back to the base images also see improvements since they can leverage the lineage information and do not need to rescan the base data.
q3 requires a backtracing query to match the bounding boxes and the OCR output in pixels on the original image and has a 41x improvement. Similarly, q6 runs 2.5x faster. q5 is illustrative of a query whose predicate does not benefit from any of the available indexes.

\begin{figure}[t]
% \vspace{-5pt}
\centering
 \includegraphics[width=0.95\columnwidth]{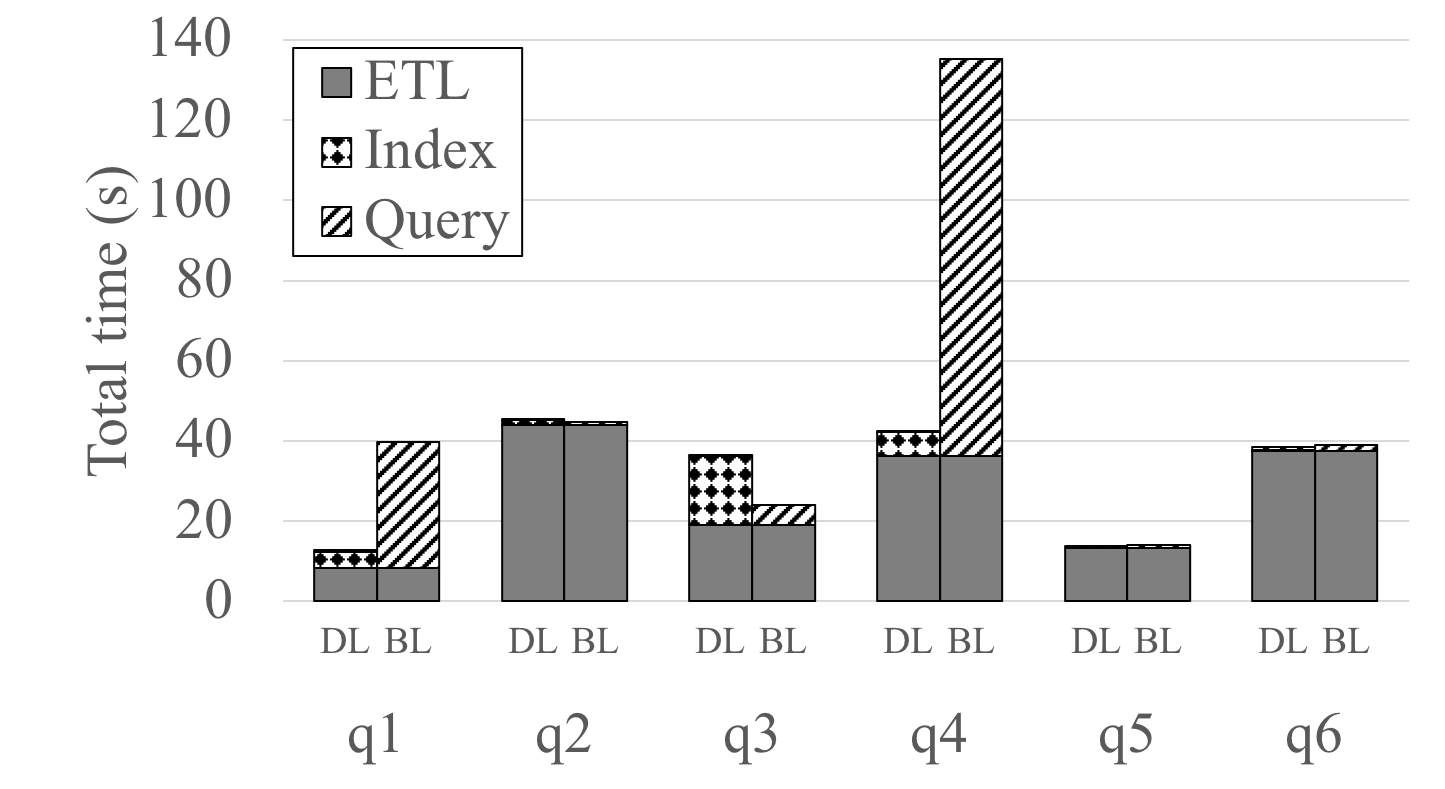}
 \caption{We evaluate the pipeline runtime, including ETL and "on-the-fly" index creation, for an optimized \textsf{DeepLens} (DL) vs. the baseline (BL). On many queries it is beneficial to materialize intermediate results and build indexes to speed up future performance. Indexing has a relatively small overhead given the compute-intensive nature of the queries.  \label{index} }
\end{figure}

\begin{figure}[t]
% \vspace{-5pt}
\centering
 \includegraphics[width=0.95\columnwidth]{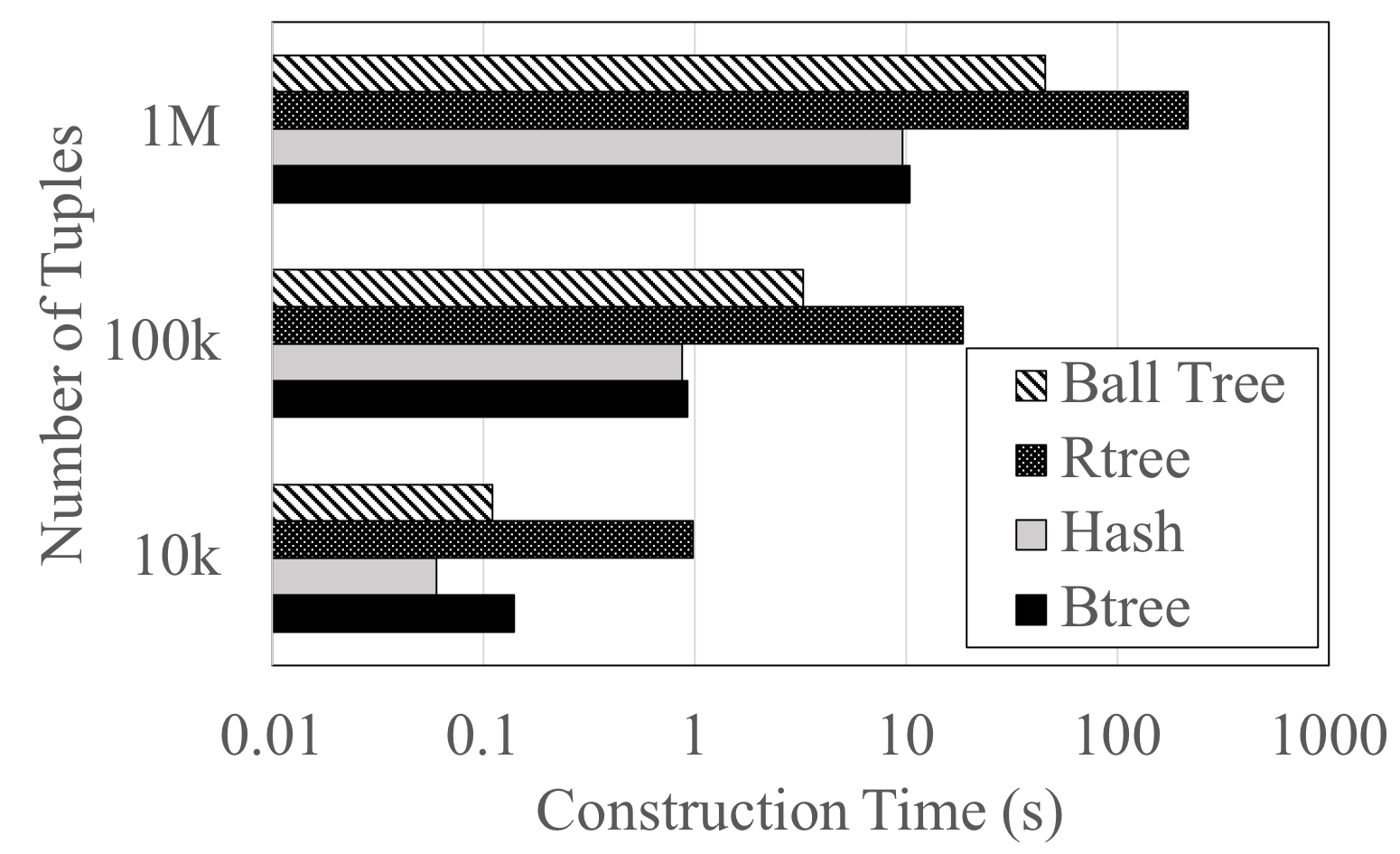}
 \caption{Building multidimensional indexes can be very costly and initial experiments indicate that construction time scales poorly with the increase of data size. \label{indexbuild} }
\end{figure}

\subsection{Overhead For Indexing}
Existing systems in this domain do not make a distinction between Query time and ETL time.
We believe this separation is justified because indexes can be reused by other queries and their construction costs amortize.
Regardless, several of the queries execute faster even if the indexes are built ``on-the-fly'' (Figure \ref{index}).
For example, q1 executes nearly 5 times faster than the baseline and q4 executes 3.5 times faster than the baseline.
The index significantly reduces the number of image matching operations that dominate the runtime, and thus, the cost of building and persisting the index is offset.

The indexes with largest overhead to build are the spatial indexes.
Figure \ref{indexbuild} plots the construction time of the multidimensional and single dimensional indexes supported in \textsf{DeepLens} as a function of the number of tuples indexed.
The R-Tree is nearly 20x slower to construct than a B+ Tree.
However, there are some mitigating factors that are interesting to consider in future work.
Since visual analytics is approximate by nature, perhaps exact multidimensional indexing is unnecessary. 
For some workloads may suffice to apply single dimensional indices and merge results with independence assumptions.
For others, locality sensitive hashing or similar approximations may suffice.

\begin{figure}[t]
% \vspace{-5pt}
\centering
 \includegraphics[width=0.95\columnwidth]{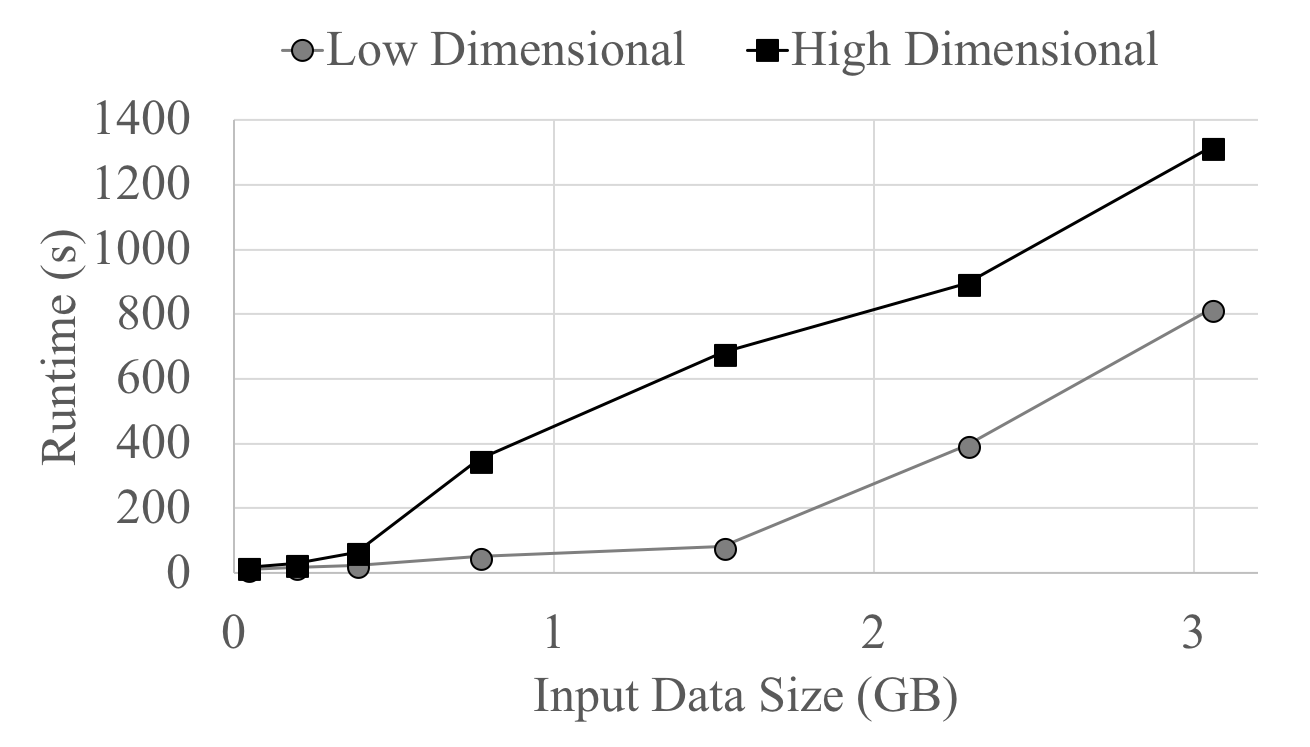}
 \caption{We evaluate the execution time of a Ball-Tree join as function of the size of the indexed relation in the high-dimensional and low-dimensional case. As the data structure is increasingly filled the execution time grows non-linearly. \label{join} }
\end{figure}

\subsection{Subtleties in Query Optimization}
For queries that join multiple image collections, neural network inference time does not necessarily dominate.
Optimizing these spatial joins turned out to be more challenging than we thought.

\subsubsection{Nonlinear join costs}
To avoid a large library of hand-written rules, the natural question to ask is whether it is possible to build a cost-based query optimizer for this system.
Figure \ref{join} illustrates the difficulties with the execution time of a Ball-Tree join as function of the size of the indexed relation in the high-dimensional and low-dimensional case. As the data structure is increasingly filled the execution time grows non-linearly. The non-linearity is also data-dependent and is more extreme in higher dimensional data. Accurately modeling the relationship between input relation size and operator cost is crucical for cost-based query optimization. Non-linearities are also known to affect standard join ordering heuristics~\cite{krishnan2018deeprljoins}.

\subsubsection{Balancing CPU vs. GPU}
\begin{figure}[t]
% \vspace{-5pt}
\centering
 \includegraphics[width=0.48\columnwidth]{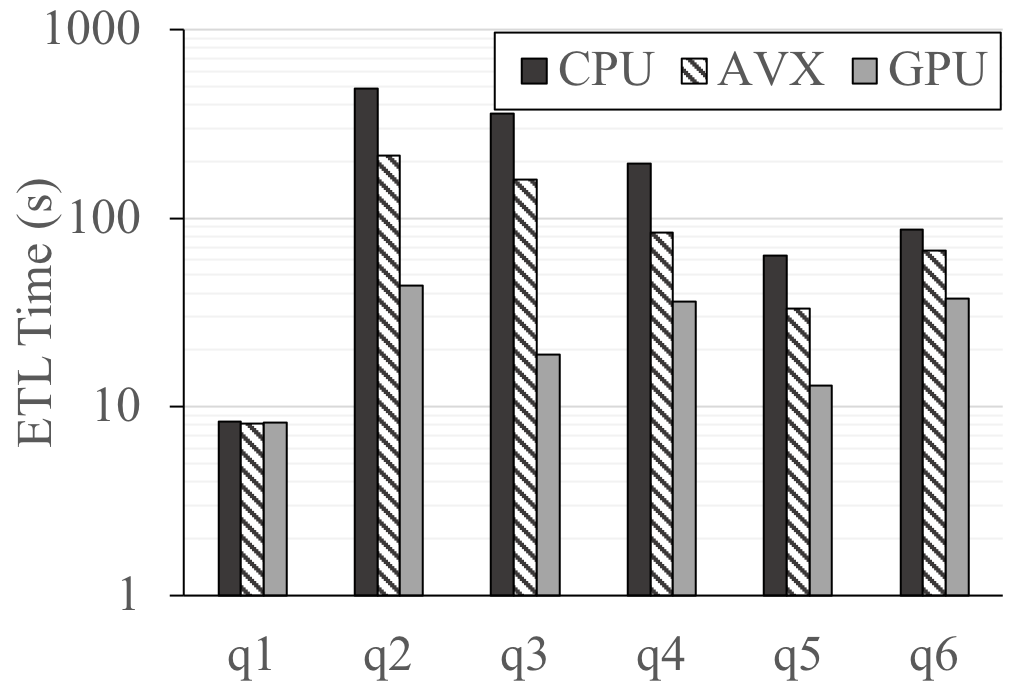}
  \includegraphics[width=0.48\columnwidth]{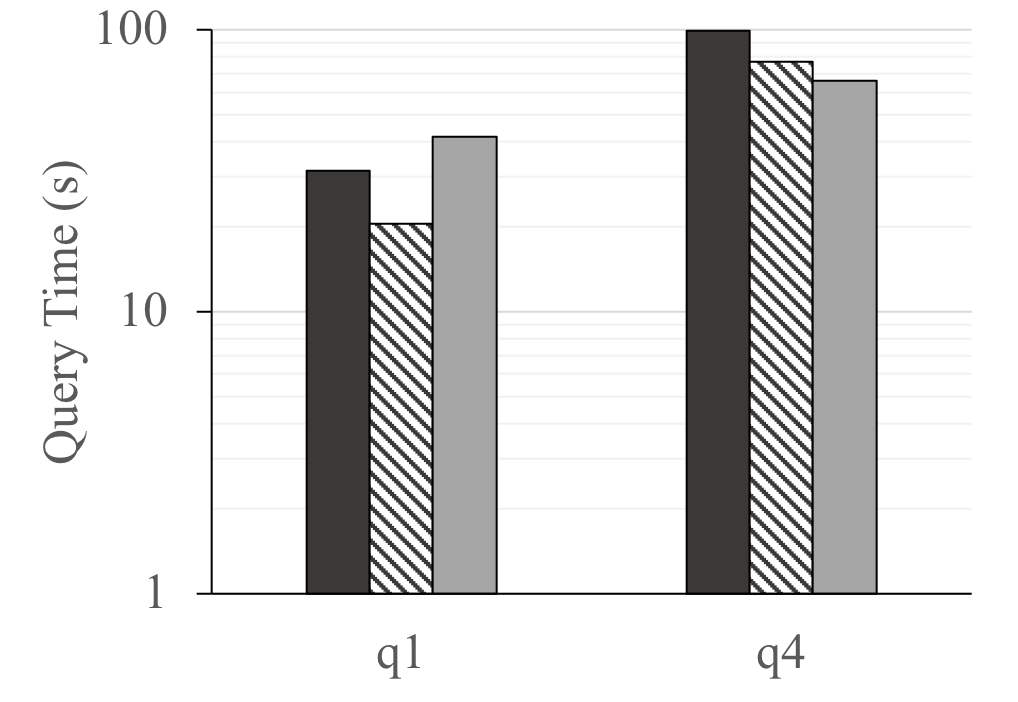}
 \caption{The execution architecture has a considerable impact on both ETL and Query time: vanilla CPU implementation (CPU), a vectorized execution (AVX), and GPU implementation (GPU). GPUs are significantly more efficient for the neural network dominated ETL time, but have more mixed results for the subsequent query processing on two image matching queries (q1, q4).  \label{build} }
\end{figure}

Processing these complex queries requires a nuanced strategy in balancing CPU vs. GPU usage. 
For the ETL phase of these queries, which is dominated by neural network inference time, using the GPU is almost universally better.
Figure \ref{build} (left) plots the processing time on each of the six benchmark queries for a vanilla CPU implementation (CPU), a vectorized execution (AVX), and a GPU implementation (GPU). Just by changing the underlying execution architecture there were up-to 12x changes in execution time. 
On the other hand, the results were more mixed for the query time. 
For the two image matching queries, q1 and q4, we implemented an all pairs matching comparison with vectorization and on the GPU.
For the larger query (q4) there is a significant performance benefit from using the GPU (34\% faster).
For the smaller query (q1), the overhead of using the GPU outweighs the costs.
Any cost-based optimizer has to weigh overheads before selecting a plan.

\subsubsection{Accuracy implications of different query plans}
Unlike in relational databases, queries in a VDMS are approximate by nature.
Cascades of approximate operators can have correlations in their errors that are hard to reason about.
These errors can be affected by query optimizer choices.
Let us take q4 as an example.
To process this query, the system first identifies patches using an object detector, then filters those patches to those that are people, and then matches all pairs of patches to deduplicate.
In another approach, the system would first identify patches using an object detector, then match all pairs of patches (regardless if they are ``people'') to deduplicate, and finally filter on those pairs that have at least one person label.
The second approach goes against typical query optimization principles of filter pushdown--but we see that it is actually a more accurate strategy.

\begin{table}[]
\centering
\begin{tabular}{|c|c|c|c|}
\hline
Execution method for q4 & Recall               & Precision & Runtime         \\
\hline
Patch, Filter, Match & 0.73      & 0.97    & 34.56  \\
Patch, Match, Filter & 0.82      & 0.98    & 62.11  \\
\hline
\end{tabular}
\caption{Accuracy vs. runtime for different execution methods of query q4.} \label{tab:q4execution}
\end{table}

\section{Conclusion and Future Work}
The growing maturity of neural networks for image classification and segmentation problems has made visual analytics an attractive and inter-discplinary field of study.
We explore the intersection of these neural network models and data management by designing a query processing engine called \textsf{DeepLens}.
Our long-term goal is develop a scalable and performant visual data management system.
While recent work has been very focused on speeding up neural network inference~\cite{kang2017noscope, zhang2018ffs, anderson2018physical, jiang2018chameleon}, we find that as we start processing increasingly complex queries, the neural network inference time no longer dominates and our experiments illustrate other bottlenecks in the system.
Addressing these complex queries requires a unified data and query model and the logical-physical separation seen in a traditional RDBMS.

Perhaps, one of the most compelling reasons to study the design of VDMS is that it brings together many hard challenges in database research. 
In particular, we believe the query optimization challenge here is significant.
We hope to explore new strategies for joint multi-query optimization and optimal physical design. Doing so might well require significant machine learning to navigate a noisy and analytically complex cost model. 
Image queries are always approximate, and managing uncertainty and quantifying the accuracy effects of a certain plan will be a serious challenge.
We plan on considering improved techniques for image compression and approximate query processing.
The natural numerical representation of images provides us with more structure than we typically have in normal database workloads.

\vspace{0.5em}
\noindent \textbf{Acknowledgements: We thank Eugene Wu and Michael Maire for their helpful feedback on early drafts. We also acknowledge NVIDIA for a generous equipment grant. This work is supported in part by the Center For Unstoppable Computing (CERES) at the University of Chicago.}

{
\fontsize{9.0pt}{11.0pt} \selectfont
\bibliographystyle{abbrv}
\bibliography{ref} 
}

\end{document}